\documentclass[intlimits,twoside,a4paper]{article}

\usepackage{graphicx}
\usepackage[T2A]{fontenc}
\usepackage[cp1251]{inputenc}

\usepackage[eqsecnum]{cmpj3}

\issue{2017}{20}{3}{33804}
\doinumber{10.5488/CMP.20.33804}

\title[Self-assembled metallic nanoparticle template]
{Self-assembled metallic nanoparticle template --- a~new approach of surface nanostructuring at nanometer scale\footnote{This work was initiated few years ago and the authors had the great opportunities to discuss the results with Jean-Pierre Badiali. We are extremely grateful for that.}}
\author[A. Taleb, V. Ivanova]{A. Taleb\refaddr{label1,label2}, V. Ivanova\refaddr{label3} }
\addresses{
\addr{label1} PSL Research University, Chimie ParisTech --- CNRS, Institut de Recherche de Chimie Paris, 75005, Paris, France
\addr{label2} Universit\'e Pierre et Marie Curie, 4 Place Jussieu, 75231 Paris Cedex, France
\addr{label3} CEA Tech, MINATEC Campus, 17 rue des Martyrs, 38054 Grenoble, France
}
\date{Received May 7, 2017, in final form July 29, 2017}

\begin{document}

\maketitle

\begin{abstract}
In the present work, the formation of silver and copper nanostructures on highly oriented pyrolytic graphite (HOPG) modified with self-assembled gold nanoparticles (Au NPs) is demonstrated. Surface patterning with nanometer resolution was achieved. Different methods such as field emission scanning electron microscopy (FEGSEM), energy dispersive spectrometry (EDS) and X-ray photoelectron spectroscopy (XPS) were used to illustrate a selective deposition of silver and copper on Au NPs. The mechanism of silver and copper ions reduction on Au NP with $n$-dodecanethiol coating is discussed. 
\keywords nanopatterning, electrodeposition, gold nanoparticles, silver and copper deposits 
\pacs 81.07.-b, 81.16.-c, 81.16.Rf, 81.15.Pq, 81.16.Dn
\end{abstract}

\section{Introduction}
Most recently, structured materials at the nanometre scale have attracted even
more attention due to the attempts to develop nanotechnological devices in many
fields such as information storage \cite{ref1,ref2}, optoelectronics \cite{ref3,ref4}, electrochemical
sensing \cite{ref5,ref6}, solar cells \cite{ref7,ref8}, catalysis \cite{ref9}, etc. The spatial periodicity of
the nanostructure is expected to play a crucial role in the determination of the
device properties \cite{ref10,ref11,ref12}. For this purpose, different strategies have been
developed to pattern the surface at the nanometer scale with high resolution and
accuracy. Among the strategies used, a self-assembled colloidal particle
template with well-defined geometry \cite{ref13,ref14,ref15} provides a simple and easy strategy
to achieve nanostructured materials. This strategy is a ``bottom up'' approach
which allows the use of a wide choice of components to enable access to diverse
physical properties with accurate control.

For the surface nanostructuring, two approaches were used: direct deposition \cite{ref16}
and deposition on a modified substrate (template) \cite{ref17}. Using a scanning tunneling
microscope, localized electrodeposition of metal has been successfully achieved, 
but it is difficult to control the location of a metal deposit with high lateral 
precision \cite{ref18}. However, a direct deposition patterning usually has the
disadvantage of achieving a disordered pattern due to the preferential 
nucleation at surface defects such as crystalline defects and step edges 
\cite{ref19,ref20,ref21,ref22,ref23,ref24}. To overcome this problem, a templated surface with patterns acting as
preferential nucleation sites is proposed. Among the techniques used for metal 
deposition, the electrochemical deposition became a versatile tool for surface
structuring and modification at nanometer scale \cite{ref19,ref20,ref21,ref22,ref23,ref24,ref25}. It is relatively cheap
and highly selective in that the new phase is deposited on the desired location.

The properties of electrodeposits depend on the surface properties, such as its 
topographical and chemical roughnesses \cite{ref26}. Dos Santos Carlo et al. show that
silver coated nanocavities act as preferred sites for silver electrodeposition 
\cite{ref26}. Another strategy for surface patterning is the creation of preferential
nucleation sites for electrodeposition at a very low scale using a mask made of 
self-assembled molecular film \cite{ref27,ref28} or colloidal particles \cite{ref29,ref30,ref31}. Most of
the published work deals with the use of the particles as a mask, and only a few 
works concern their use as nucleation sites. A recently published method  
that is particularly relevant to this work showed vertical ZnO nanowire arrays 
growth on an Au NPs-modified ZnO seed layer by using the vapor-liquid-solid
(VLS) deposition method \cite{ref29}. Metallic particles act as templates, through
appropriate curvatures and/or their metallic nature, to create preferential 
deposition sites on the surface.

In this work, the advantage of using a self-assembled Au NPs template to create
preferential sites for metal electrodeposition is reported. For a short 
electrodeposition time, there is illustrated a capability of self-assembled Au NPs
template to control metal electrodeposition at nanometer scale. Ordered 
nanostructures with nanometer resolution were prepared. It is shown that this 
patterning method allows to achieve a lower resolution of about 2~nm.

\section{Experimental details}
Silver and copper electrodeposition is performed using a three electrode electrochemical 
configuration. The reference and counter electrodes were, respectively, Ag/AgCl 
and a platinum sheet. A clean (0001) highly oriented pyrolytic graphite (HOPG) 
surface was used as a working electrode and was prepared through a cleaving process. 
The modified HOPG electrode is prepared by depositing a droplet of Au NPs
solution on the HOPG surface. Silver is deposited from an aqueous solution of
$10^{-2}$~M AgNO$_3$ and 1~M HClO$_4$ as a supporting electrolyte. Cu was deposited from
the aqueous solution of $10^{-3}$~M CuSO$_4$ and 1~M H$_2$SO$_4$ as a supporting
electrolyte. These solutions are deaerated by purified nitrogen during 2~h
before use. AgNO$_3$, HClO$_4$ are obtained from Fluka. CuSO$_4\cdot\text{5H}_2$O and H$_2$SO$_4$
are purchased from Merck. All the chemicals used are of analytical grade and are used without 
further purification. The electrolytes are prepared with water purified by Milli 
Q system (Millipore, electric resistivity 18.2~M$\Omega\cdot\text{cm}$).

The 4~nm Au and silver (Ag) NPs are synthesized using the Stucky method \cite{ref32}.
After the synthesis, Au and Ag NPs are coated with $n$-dodecanethiol and present a
narrow size distribution (6\%). By depositing a droplet of Au or Ag NPs solution
on HOPG electrode, a 2D assembly of these nanoparticles in a hexagonal structure
is obtained.  

Electrochemical measurements (chronoamperometry) are performed with a
Voltalab PGZ301 potentiostat and are carried out at room temperature. The
chronoamperometry is obtained at constant potentials of $-0.58$~V (vs. SCE) and
$-0.3$~V (vs. Ag/AgCl), respectively, for copper and silver electrodeposition over
10~s. 

The morphologies of silver and copper electrodeposits and Au NPs film are investigated,
using a high-resolution Ultra 55 Zeiss field emission gun scanning electron
microscope (FEGSEM) and a transmission electron microscope (JOEL 100CX) operated
at 100~kV.

Chemical compositions of the deposits are identified in FEGSEM using a PGT
spirit energy dispersive spectrometry system (EDS). Additionally, for chemical
composition determination, we used X-ray photoelectron spectroscopy (XPS)
realized with a thermo VG Scientific ESCALAB 250 system fitted with a
microfocused, monochromatic AL K$\alpha$ X-ray source ($h\nu = 1486.6$~eV, spot size 650~{\textmu}m, $\text{power} = 15~\text{kV} \times 200$~W).

\section{Results and discussion}
\begin{figure}[!t]
\begin{center}
\includegraphics[width=0.7\textwidth]{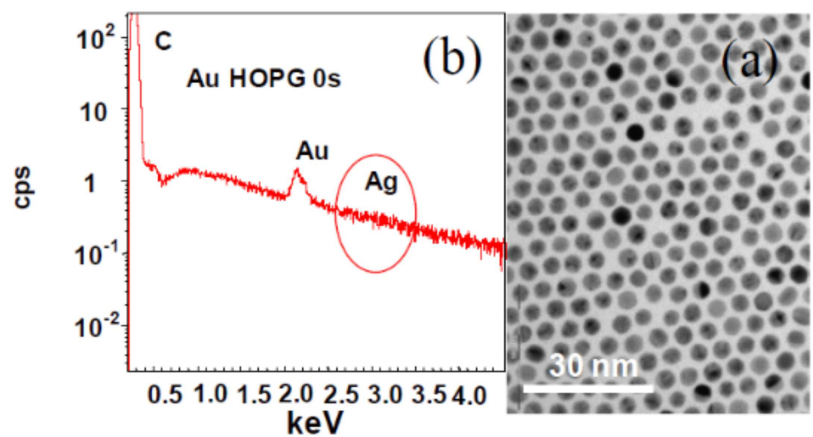}
\caption{(Color online) (a) TEM image of self-assembled Au NPs in a 2D hexagonal structure. (b)
EDS spectrum taken for Au NPs-modified HOPG electrode. }
\label{fig1}
\end{center}
\end{figure}

The HOPG electrode was modified by drop coating using a droplet of Au NPs
solution on the surface, which then self-organizes in a hexagonal structure 
[figure~\ref{fig1}~(a)]. It has been shown that the nanoparticle size, shape and surfactant 
shell are the main parameters determining the structure formed in the 
self-assembly process \cite{ref33}. Thus, for ordered closely packed nanoparticles arrays
with long range, both a narrow size distribution and an adequate capping ligand 
surface coverage are required. In fact, the passivating ligand lubricates the 
interface between the nanoparticles and the HOPG surface enhancing the 
nanoparticles diffusion to form a well arrangement in long range. Moreover, 
this shell should be highly repulsive at a close range but thin enough to maintain 
a short interparticle separation, a crucial factor in the physical properties 
(electronic and optical) of metal nanoparticles assembly. In our case, the 
optimum conditions for the formation of a close 2D hexagonal array with a high 
degree and long range ordering observing the arguments mentioned above are 
obtained using $n$-dodecanethiol molecules. The long range ordering is crucial for a
surface nanostructuring approach proposed in the present work.

The structural parameters of the organized array of $n$-dodecanethiol-capped Au
NPs depend on the configuration of the $n$-dodecanethiol alkyl chain. This 
surfactant serves not only as a protective layer for the nanoparticles, preventing 
the coalescence and the consequent collapse of the structure, but also fixes the 
interparticle distance \cite{ref13}. The distance between two nanoparticles obtained
from TEM patterns is found to be between 1.8 and 2~nm. This similarity between 
the average interparticle distance and the length of alkyl chains suggests the 
intercalation or interpenetration of individual chains or domains of chains~\cite{ref34}.

\begin{figure}[!t]
\begin{center}
\includegraphics[width=0.69\textwidth]{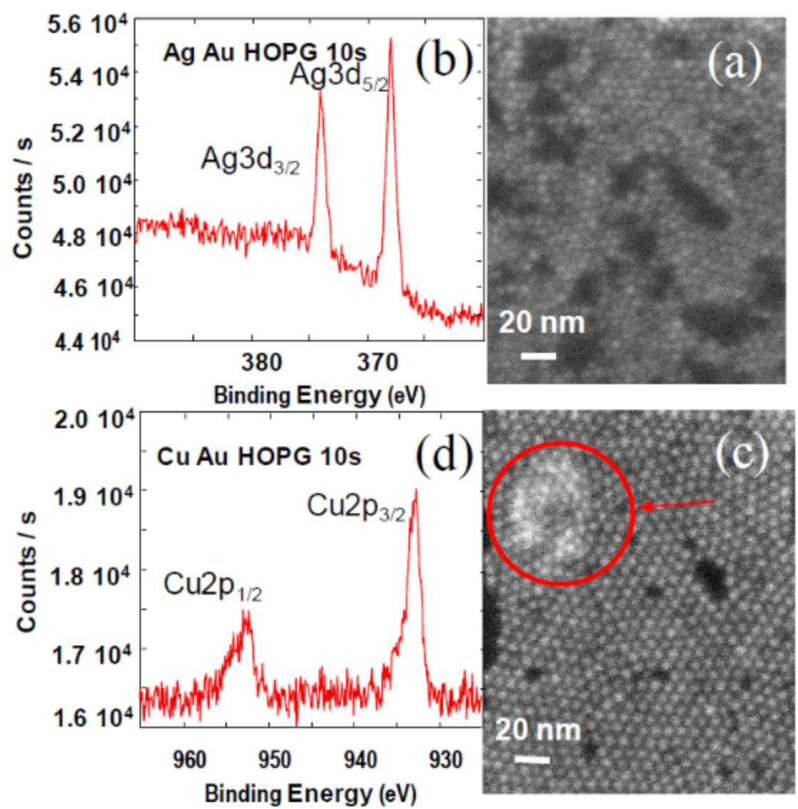}
\caption{(Color online) (a) FEGSEM images of silver electrodeposits onto Au NP-modified HOPG
electrode from an aqueous solution of $10^{-2}$~M AgNO$_3$ and 1~M HClO$_4$ at potential
deposition of $-0.3$~V for the deposition times of 10~s. (b) High resolution XPS
spectra of Ag3$d$ core level corresponding to Ag deposits. (c) FEGSEM images of Cu
deposits obtained by electrodeposition from a solution of $10^{-3}$~M CuSO$_4$ and 1~M H$_2$SO$_4$
on Au NPs-modified HOPG electrode over 10~s and at the applied potential of $-581$~mV vs. SCE. (d) High resolution XPS spectra of Cu2$p$ core level corresponding to Cu deposits.}
\label{fig2}
\end{center}
\end{figure}

FEGSEM images shown in figure~\ref{fig2}~(a) and \ref{fig2}~(c) are the obtained results of Ag and Cu
electrodeposition achieved with an Au NPs modified HOPG electrode. It should be
noted that Cu and Ag deposits have taken place on Au NPs and appear as  bright
spots. Furthermore, homogeneous distribution of Cu and Ag deposits on the surface
can be observed.

\begin{figure}[!t]
\begin{center}
\includegraphics[width=0.485\textwidth]{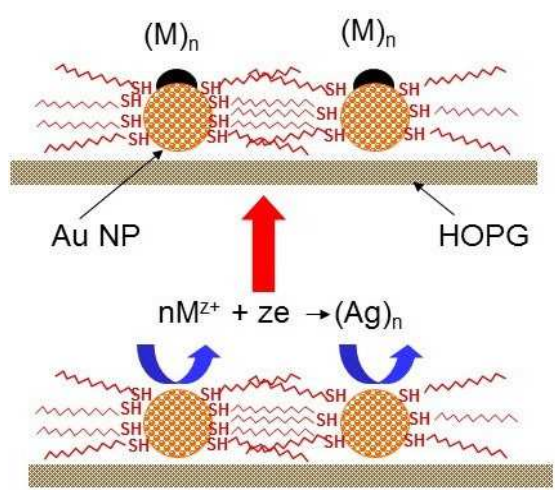}
\caption{(Color online) Schematic illustration of metal (Ag and Cu) electrodeposition on Au NPs
monolayer, showing how the dense bundles of $n$-dodecanethiol ligands between Au NPs act as a mask.}
\label{fig3}
\end{center}
\end{figure}

The EDS and XPS analyses have been performed to determine the chemical
composition of the electrodeposited nanoparticles. Figure~\ref{fig1}~(b) represents the EDS
spectra of Au NPs-modified HOPG electrode before metal electrodeposition (Ag and
Cu). It should be noted that only Au signal is observed, corresponding to Au NPs
modified HOPG substrate. After 10~s copper and silver electrodeposition, this should be noted by a
high resolution XPS spectrum, the appearance of different peaks corresponding to Ag3$d$ and Cu2$p$ core
level [figure~\ref{fig2}~(b) and \ref{fig2}~(d)] appeared. 

For a further illustration of the patterning effect of Au NPs, FEGSEM patterns are
compared to TEM image of Au NP-modified HOPG electrodes before metal
electrodeposition. It can be seen that after 10~s electrodeposition, the
hexagonal pattern formed by Au NPs is still preserved and the Ag and Cu deposits
are reproducing this initial structure [figure~\ref{fig2}~(a) and \ref{fig2}~(c)]. These results
exclude any silver or copper electrodeposition between the Au NPs, otherwise
coalesced nanoparticles should be observed. In the indicated region of figure~\ref{fig2}~(c) there can be observed
Cu nanoparticles bigger than Au NPs diameter (4~nm). It is
also seen that some Cu nanoparticles start to coalesce with the disappearance of
the hexagonal structure of Au NPs assembly.

These results clearly show that using an electrodeposition process and Au NPs
template a surface patterning with a resolution, deposit-to-deposit distance, of
around 2~nm is achieved. This resolution is limited by the Au NPs shell-coating
(figure~\ref{fig3}). For comparison, the electron beam lithography, allows surface
patterning with a resolution of about 10~nm \cite{ref35}, where the resolution is
controlled by the electron beam size, and by the forward electron scattering.
Furthermore, for optical lithography, the resolution is also about 20~nm and it
is restricted by the diffraction limit, with the advantage of being a more scalable
technique to a large area of about a few square centimetres \cite{ref36}. By contrast,
electron lithography has low throughput and covers small areas.

Furthermore, compared to colloidal lithography in which colloidal particles are
used as a mask for surface nanostructuring \cite{ref37}, our method has an advantage of
using metallic nanoparticles both as a mask and a template for metal
electrodeposition. In addition, it offers an alternative and cheaper method of
producing a quite regular pattern of metal nanoparticles over a long range surface.
On the other hand, the patterning resolution is limited in the case of colloidal
lithography by the nanoparticles size, whereas in the present case it is limited
by the coating of Au NPs, which fixes the distance between them.

The results reported here show that Au NPs could be used as a template for
further electrochemical deposition and thus nanostructuring the surfaces, which
could be used for others materials. We believe that the use of metallic
nanoparticle templates for nanostructuring of different surfaces opens up new
opportunities for different nanotechnological applications.

\section{Conclusions}
In the present work, we have shown that Au NPs-modified HOPG electrode could be
used as a template for the fabrication of silver and copper nanostructures by
electrochemical deposition. The resulting silver and copper deposits perfectly
reproduce the hexagonal network of self-organized Au NP patterns in a nanometer
resolution of about 2~nm, confirming a preferential growth on the Au NPs. The
obtained results show the use of self-assembled metallic nanoparticles as
an interesting way to achieve a surface patterning at nanometer scale.

\section*{Acknowledgements}
We are grateful to Pierre Dubot at the Institut de chimie et des mat\'eriaux Paris
Est UMR 7181 for acquiring and discussing the XPS spectra.

\ukrainianpart
\title{Самоорганізований шаблон металічних наночастинок --- новий метод наноструктуризації поверхні у нанометровому масштабі
}
\author{A. Талеб\refaddr{label1,label2}, В. Іванова\refaddr{label3} }
\addresses{
\addr{label1}  Дослідницький університет науки та літератури Парижу, Chimie ParisTech --- CNRS, Інститут хімічних досліджень Парижу, Париж, Франція
\addr{label2} Університет П'єра і Марії Кюрі,  Париж, Франція
\addr{label3} CEA Tech, MINATEC Кампус,  Гренобль, Франція
}

\makeukrtitle

\begin{abstract}
В даній роботі продемонстровано наноструктури срібла і міді на високо орієнтованому  піролітичному графіті (HOPG), 
модифікованому самозгромадженими золотими наночастинками  (Au NPs). Отримано формування рисунку на поверхні з 
нанометровою роздільною здатністю. Різні методи, такі як автоелектронний скануючий  мікроскоп (FEGSEM), енергетично 
розсіювальний спектрометр  (EDS) і X-променева фотоелектронна спектроскопія, були використані для того, щоб проілюструвати  
селективне осадження срібла і міді на Au NP. Обговорено механізм меншання іонів срібла і міді на  Au NP з покриттям n-додеканетіолом.
\keywords наноструктурування, електроосадження, золоті наночастинки, осади золота і міді
\end{abstract}

\end{document}